\documentstyle[12pt,aasms4]{article}   

\begin{document}
\title
{WATER ICE, SILICATE AND PAH EMISSION FEATURES
IN THE {\em ISO} SPECTRUM OF THE CARBON-RICH
PLANETARY NEBULA CPD--56$^\circ$8032\footnote{Based on observations
with {\em ISO}, an ESA project with instruments funded by ESA Member
States (especially the PI countries: France, Germany, the Netherlands and
the United Kingdom) with the participation of ISAS and NASA}}

\author
{M{\sc artin} C{\sc ohen}$^1$, M. J.
B{\sc arlow}$^2$, R. J. S{\sc ylvester}$^2$, X.-W. L{\sc iu}$^2$,
P. C{\sc ox}$^3$, T. L{\sc im}$^4$, B. S{\sc chmitt}$^5$, 
A. K. S{\sc peck}$^2$\\
$^1$Radio Astronomy Laboratory, 601 Campbell Hall, University of 
California, Berkeley, CA 94720; email: mcohen@astro.berkeley.edu\\
$^2$Dept. of Physics and Astronomy, University College London,
Gower Street, London WC1E 6BT, U.K.\\
$^3$Institut d'Astrophysique Spatiale, B\^atiment
121, Universit\'e de Paris XI, F-91405 Orsay Cedex, France\\
$^4$The LWS-Instrument--Dedicated Team, {\em ISO} Science
Operations Centre, P.O. Box 50727, E-28080 Madrid, Spain\\
$^5$Laboratoire de Glaciologie et Geophysique de 
l'Environnement, CNRS, 54 rue Moliere, BP 96, F-38041 Grenoble/Saint
Martin d'Heres, France}

\flushbottom

\begin{abstract}
Combined ISO SWS and LWS spectroscopy is presented of the late WC-type 
planetary nebula nucleus CPD-56$^\circ$8032 and its carbon-rich nebula.
The extremely
broad coverage (2.4--197~$\mu$m) enables us to recognize the clear
and simultaneous presence of emission features from both 
oxygen- and carbon-rich circumstellar materials.   Removing a
smooth continuum highlights bright emission bands characteristic of
polycyclic aromatic hydrocarbons (hereafter PAHs) in the 3--14-$\mu$m
region, bands from crystalline silicates longwards of 18~$\mu$m, and the
43- and 62-$\mu$m bands of crystalline water ice. We discuss the
probable evolutionary state and history of this unusual
object in terms of (a) a recent transition from an O-rich to a
C-rich outflow following a helium shell flash; or (b) a carbon-rich
nebular outflow encountering an O-rich comet cloud.
\end{abstract}


\section{Introduction}

CPD--56$^\circ$8032 (hereafter CPD) belongs to the rare class of late
WC-type nuclei of planetary nebulae and is classified as [WC10] in the scheme
of Crowther, De Marco \& Barlow (1998). It is thought that such objects may
be one result of helium shell-flashes in low- and intermediate-mass stars
on the Asymptotic Giant Branch (AGB).  For a certain fraction of these 
double-shell burning stars, a helium shell flash may have
ejected or ingested essentially all the remaining hydrogen-rich outer
envelope.
The resulting star could then be H-poor, like the late WC-type ([WCL]) 
planetary nebula nuclei (PNNs) whose spectra essentially mimic those of
{\it bona fide} population I Wolf-Rayets, although mostly with lower wind
velocities.  

The large near- and mid-infrared excess of CPD has been known for over
twenty years (Webster \& Glass 1974; Cohen \& Barlow 1980; Aitken et al. 
1980), and has been attributed to emission by dust grains.  Mid-infrared
emission bands, most often attributed to PAHs (e.g., Allamandola,
Tielens, \& Barker 1989) were detected in
ground-based 8--13-$\mu$m spectra by Aitken et al. (1980) and in airborne
5--8-$\mu$m spectra by Cohen et al. (1989). Longer wavelength PAH
bending modes were identified by Cohen, Tielens \& Allamandola (1985) 
from 7.7--22.7 $\mu$m spectroscopy of CPD with the IRAS Low
Resolution Spectrometer (hereafter LRS). 
CPD has the highest measured luminosity fraction in the 7.7-$\mu$m band
of any object (Cohen et al. 1989)
and its nebula is characterized by a gas-phase C/O number ratio of 13
(De Marco, Barlow \& Storey 1997; hereafter DMBS97), the joint
highest gas-phase C/O ratio measured for a planetary nebula. 

We present Infrared Space Observatory (ISO) Long Wavelength Spectrometer
(LWS;  Clegg et al. 1996; Swinward et al. 1996) 43--197-$\mu$m full
grating spectra of CPD, combined with Short Wavelength Spectrometer (SWS;
de Graauw et al. 1996) 2.4--45-$\mu$m grating spectra of this object,
obtained in the LWS Guaranteed Time program. 
Preliminary results on the ISO spectra of CPD were presented by Barlow
(1998).

\section{The ISO spectrum of CPD-56$^{\rm o}$8032}

Full wavelength coverage grating mode LWS01 spectra of CPD were secured
during ISO revolution 84. The spectral resolution was 0.6-$\mu$m in first
order (84--197-$\mu$m) and 0.3-$\mu$m in second order (43--93-$\mu$m). The
spectra consisted of eight fast scans, each comprising a 0.5-s
integration ramp at each grating position, sampled at 1/4 of a spectral
resolution element.  Our low-resolution 2.4--45-$\mu$m SWS grating
spectrum of CPD was taken during ISO revolution 273. The SWS01 AOT was
used at Speed 1, yielding a mean spectral resolving power of 200--300
over the whole spectrum.  
Standard pipeline processing (LWS OLP6.0 and SWS OLP6.1) was used to
extract, reduce and calibrate all the separate spectral fragments. The
ISAP and SIA packages provided the capability to examine the spectral
fragments in detail.

Fig.~1 presents our complete wavelength coverage of CPD, combining SWS and
LWS data.  First we spliced all the SWS and LWS subspectra separately, 
following the methods described by Cohen, Walker \& Witteborn (1992; CWW),
then joined the composite SWS and LWS portions, which required scaling the
LWS spectrum by 0.99$\pm$0.01 to register it with the SWS spectrum.  The
resultant 2.4--197-$\mu$m spectrum was normalized to the Point Source
Catalog (PSC) photometry (see CWW) in all 4 IRAS bands.  This
necessitated a further rescaling of the total spectrum of CPD by a factor
1.05$\pm$0.04.  


38--90-$\mu$m (lower).  

The ISO spectrum of CPD in Fig.~1 exhibits unresolved emission lines of
[C~${\sc ii}$] 158-$\mu$m, [O~${\sc i}$] 63- and 146-$\mu$m and CO
rotational lines between J~=~14--13 at 186.0~$\mu$m and J~=~19--18 at
137.2~$\mu$m, which are all excited in the nebular photodissociation
region. The canonical spectrum of PAH emission bands dominates the peak of
Fig.~1 below 15~$\mu$m but the most striking aspect is that, despite the
carbon-dominated stellar and nebular chemistry, the spectrum longwards of
15~$\mu$m is dominated by emission features usually associated with the
circumstellar envelopes of O-rich stars (Glaccum 1995, Waters et al. 1996,
Waelkens et al. 1996). Waters et al. (1998) have presented SWS spectra
of two other C-rich PNe with [WCL] nuclei, which also show both PAH and
crystalline silicate emission features. The ISO spectrum of CPD shown here
has an additional remarkable property in that crystalline water ice
features are present in emission at 43 and 62~$\mu$m (see below).

To identify and quantify the intensities of the many apparent emission
bands, we have removed all obvious features from the SWS+LWS spectrum to
define a set of continuum points longwards of about 4.5~$\mu$m.  The
simplest fit to this continuum was the sum of two blackbodies, with their
temperatures and solid angles optimized by least-squares fitting to
provide a lower envelope to the observed spectrum. This lower bound was 
achieved by ensuring that the difference spectrum (observed-minus-continuum)
is negative only over small wavelength intervals in order to avoid
truncating any emission features. The best fit is
achieved for temperatures of 470$\pm$5K and 135$\pm$5K, which we interpret
as grains within the ionized nebula heated by direct starlight (470K) or
by resonantly trapped Lyman$\alpha$ photons.  Subtraction of
this simple continuum yields the CPD ``excess'' spectrum. Below 5~$\mu$m, 
the steep excess can be attributed to hot grains ($\sim$1600K). It is far 
above the stellar continuum radiation
calculated by De Marco \& Crowther (1998) and reddened using DBMS97's
value of E(B-V)=0.68, which contributes no more than 5\% of the emission
at 2.4~$\mu$m and rapidly diminishes with increasing wavelength. 

The standard PAH emission bands (Cohen et al. 1989) dominate the excess
(Fig.~1) with
features at 3.3, 5.2, 6.2, 6.9, 7.7, 8.7 and 11.3~$\mu$m, together with
the underlying emission plateaus at 6--9~$\mu$m and 11--14~$\mu$m. Between
16 and 40~$\mu$m (Fig.~2a), several emission features characteristic of
crystalline silicates (Glaccum 1995, Waters et al. 1996) are recognizable,
most prominently at 19, 24, 28,
and 33~$\mu$m. The longest wavelength region (Fig.~2b)  exhibits emission
bands at 41, 43, 47.5, and 69~$\mu$m, along with a very broad, low-level,
emission hump centered near 62~$\mu$m, under the [O~${\sc i}$] 63-$\mu$m
line. We identify the 43- and 62-$\mu$m bands with crystalline water ice,
as first detected by Omont et al. (1990) in the KAO spectrum of the Frosty
Leo nebula.

Following Glaccum (1995) and Waters et al. (1996), we have attempted to
identify the emission features seen above the continuum of CPD using (a)
data for clinopyroxene, orthopyroxene and 100\% forsterite (Koike et al.
1993); and (b)  optical constants for both amorphous and crystalline water
ice recently measured in the laboratory (Trotta 1996, Schmitt et al.
1998). We chose the 100\% pure forsterite because of its weak but definite
feature near 69~$\mu$m, shown (according to Koike et al. 1993) only by the
pure form of this material.  In Fig.~2 we distinguish the separate
contributions of these four materials, and their total.  We modeled the
optically thin case, for small spherical grains (0.1 or 1.0~$\mu$m
radius), so that we could neglect scattering and represent Q$_{ext}$ by
Q$_{abs}$.  By trying to match the relative band strengths, we found
plausible temperatures for each component.  We made no attempt to
constrain these separate temperatures, but three were found to be
identical (forsterite, clinopyroxene, and crystalline ice: 65~K)  and the
orthopyroxene rather similar (90~K), perhaps suggestive of a common
physical location, or of core-mantle grains.  At these temperatures, the
cold silicates produce no measurable emission below the 19-$\mu$m band. 
Note that without combined SWS+LWS coverage, we could not constrain these
temperatures. 

Our dust modeling indicates that forsterite causes the 24- and
33-$\mu$m bands, and the weak ``wing" at 36~$\mu$m.  To match the
28-$\mu$m band requires crystalline silicates such as orthopyroxene, which
also contributes the 19-$\mu$m bending mode.  Crystalline ice produces
features near 43- and 62-$\mu$m, due to the transverse optical and
longtitudinal acoustic vibrational branches respectively.  As noted by
Omont et al. (1990), the 62-$\mu$m band is not shown by amorphous ice (see
the laboratory spectra of Smith et al. 1994), so the observation of this
band demands the presence of crystalline ice. 
The prominent feature near 41~$\mu$m is probably dominated by
clinopyroxene and this material also contributes a
broad emission feature centered near 66-$\mu$m that merges with the longer
wavelength ice band. 

The sum of our separate component emissions provides only a qualitative
match to CPD's spectrum (Fig.~2), but is surely indicative of the
circumstellar materials present, and confirms the striking
presence of oxygen-rich materials around a carbon-rich PN. 
Typically, interband and wing emissions 
associated with the laboratory features in Fig.~2a,b contribute $\sim5\%$
of the peak emission.  Our lower envelope has not removed such a large
continuum so it appears from Fig.~2 that laboratory ``crystalline" materials
yield features much too broad compared with CPD's features.  Before drawing 
this conclusion we would prefer to incorporate these materials into a 
physically realistic radiative transfer code, including grain geometry 
and secondary aspects like grain mantle structure and porosity.
Note that none of the materials we used to fit the spectrum
has a feature matching the one observed at 47.5~$\mu$m (Fig.~2b).

\section{Discussion} 

Using current estimates for the luminosity and distance of CPD, we can
deduce the angular extent of each of the emitting dust components around
it.  Adopting a distance of 1.53~kpc (DMBS97; De Marco \& Crowther 1998),
implies that, if they emit as equilibrium blackbodies, the 470~K and
135~K grains we use to represent our dust continuum must lie
36 ~AU (0.02~arcsec)  and 400~AU (0.26~arcsec) from the star,
respectively.  We have derived the distance from the star of
the crystalline silicates using optical constants in the UV and
visible by Scott \& Duley (1996).  These materials are 
transparent in portions of the short wavelength spectrum but 
highly absorbing near the UV peak of the energy distribution 
of the WCL nucleus of this nebula. A realistic energy balance between 
the UV grain absorption at this peak (we used the NLTE model atmosphere of
De Marco \& Crowther 1998) and IR re-emission
indicates that the 65--90~K oxygen-rich constituents must lie
1000~AU (0.005~pc) away, i.e., 0.60~arcsec.  
Thus even the warmest silicates and ices lie near the periphery of
the ionized nebula, based on the HST images (DMBS97) which show
all the nebular H$\beta$ emission to be confined to 
1.6$\times$2.1~arcsec. 
These dimensions, and the normalization factors involved when we fit the
emission spectra of grains to the excess emission in CPD, also yield 
rough estimates of the mass in the crystalline components.  We obtain
about 1.6, 1.3, 0.3, and 0.6 $\times10^{-4}M_\odot$ for forsterite,
clinopyroxene, orthopyroxene, and water ice, respectively, independent of
grain size, for grain radii $<3~\mu$m.

Roche, Allen \& Bailey (1986) found the 3.3-$\mu$m PAH emission to have 
an angular size of 1.3~arcsec, i.e., within the
ionized boundary of the nebula, while our postulated 470~K and 135~K
blackbody grains lie well within the
carbon-rich nebular ionized gas, suggesting that they too are likely to be
carbon-rich (though we cannot prove this). The very hot dust responsible for 
the excess continuum below 5~$\mu$m must be only a few AU from the star and
may be condensing in the wind of the WC10 central star,
like the dust found to form in the outflows from Population-I WC9 stars
(e.g., Cohen, Barlow \& Kuhi 1975). Due to the absence of hydrogen in the
wind, PAHs cannot be present there; the condensation of hydrogen-deficient
soots must occur by pathways that bypass acetylenic chains and emphasize
grain formation via fullerenes and ``curling" of graphite sheets (e.g.,
Curl and Smalley 1988).
Once such grains later penetrate into the H-rich nebula, partially
hydrogenated PAHs may be created. Some nebular PAHs may also have been
created previously in a C-rich AGB outflow, before it became completely
hydrogen-depleted. 

CPD is not the first PN to have the signatures of both
C-rich and O-rich material recognized.  IRAS~07027--7934, found by
Menzies \& Wolstencroft (1990) to be a low-excitation planetary nebula
with a C-rich [WCL] central star (WC10 in the scheme of Crowther et
al. 1998), exhibits PAH features in its IRAS LRS
spectrum. Yet Zijlstra et al. (1991)  discovered it hosts a strong
1612~MHz OH maser, normally associated only with O-rich material. The
Type~I bipolar PN NGC~6302 exhibits weak 8.7- and 11.3-$\mu$m PAH bands
(Roche \& Aitken 1986), a weak OH maser (Payne, Phillips \& Terzian 1988),
and silicate features at 19~$\mu$m (Barlow 1993) and longer wavelengths
(Waters et al. 1996). Its LWS spectrum exhibits prominent crystalline
water ice emission bands (Barlow 1998, Lim et al. in preparation). 
Waters et al. (1998) have detected crystalline silicate emission
features in the SWS spectra of the strongly-PAH emitting C-rich objects
BD+30$^{\rm o}$3639 and He~2-113 (both PNe with cool WC nuclei).

Amongst hypotheses to explain the simultaneous existence of C- and
O-rich particles around C-rich stars, two 
(Little-Marenin 1986; Willems \& de Jong 1986) are of possible
relevance to CPD and similar nebulae: (a) a recent thermal pulse has
converted an O-rich outflow to one that is C-rich; (b) the silicate
grains are in orbit around the system and existed well before the current
evolutionary phase.

If a recent thermal pulse converted an O-rich mass loss outflow to
a C-rich one, the O-rich grains should be further out and cooler than the
C-rich particles, in agreement with the properties of the dust around CPD.
The chief objection raised to this scenario, in the context
of carbon stars showing warm silicate emission, has been that such a
transition should occur only once during the lifetime of a star, so
the probability of finding a carbon star with silicate grains in its
outflow, still sufficiently close and warm to exhibit a 10-$\mu$m
feature, should be extremely small.  This objection is weakened for
extended nebulae, since the cooler nebular particles probe a longer
look-back time than do 10-$\mu$m emitting silicate grains in a carbon star
outflow.  
The expansion velocity of CPD's nebula is 30~km~s$^{-1}$ (DMBS97), so that
nebular material now 1000~AU from the star, the deduced location of the
O-rich grains, must have been ejected only 160~years ago. Even with
a lower ($\sim$10~km~s$^{-1}$) typical AGB outflow velocity, the timescale
is small
compared to the predicted interval of 6$\times10^4$~yrs between successive
thermal pulses for a 0.62~M$_\odot$ core (Boothroyd \& Sackmann 1988).
Thus a sufficiently recent O-rich to C-rich transition by CPD (as well as
by He~2-113 and BD+30) appears statistically improbable, unless, as
suggested by Waters et al. (1998), such stars are somehow particularly
susceptible to a thermal pulse during their immediate post-AGB phase. 

Because of the above difficulty, Lloyd-Evans (1990) and Barnbaum et al.
(1991) proposed that carbon stars showing silicate emission are binaries
containing a disk in which O-rich grains have accumulated during an
earlier evolutionary phase, with the extended disk lifetimes allowing
silicate emission to persist. For silicate grains to be warm enough to
exhibit a 10-$\mu$m emission feature, the inner edge of such a disk could
be at no more than 4--5 stellar radii from a carbon star (Barnbaum et
al.), i.e., 5--6~AU for a 3000~L$_\odot$, 2400~K star. An O-rich dust disk
with these parameters cannot be present around CPD, since it shows no
trace of 10-$\mu$m silicate emission.  At a radius of $\sim$1000~AU, the
cool (65--90~K) O-rich grains would not be in a conventional mass-transfer
circumstellar disk around one component of a wide binary, although Fabian
\& Hansen (1979) have suggested a mechanism whereby a wide binary system
might focus matter from one component into a helical trajectory.  An
alternative possibility that would allow the pre-existing grains
hypothesis to be retained would be if they resided in a Kuiper belt or
inner Oort comet cloud around the star. A radius of 1000~AU is comparable
to current estimates for the outer edge of the Kuiper belt around our own
Sun (Weissman 1995) -- the interaction of cometary nuclei in such a belt
with CPD's mass outflow and ionization front might provide the conditions
needed to liberate the small particles ($<3-10~\mu$m radius) that are
required in order to explain the observed far-infrared silicate and ice
bands. The annealing and recrystallization of silicates and ice grains
liberated from comets, leading to the required highly ordered structures
with correspondingly ``sharp" emission features, could result from the
sudden increase in the UV photon flux from CPD during its post-AGB
evolution. Difficulties for the comet-cloud hypothesis include (a) the
relatively large mass ($\sim$130 Earth masses) of crystalline silicates
derived for CPD, high compared to current, though still rather uncertain,
estimates for the mass of the Solar System Kuiper Belt and Oort Cloud.
However, the more massive progenitor stars of current PNe could
have appreciably higher mass comet systems; (b) the apparent correlation
between the presence of crystalline silicate grains around PNe and the
presence of both PAHs and a WCL central star seems to implicate a 
chemistry change between an O-rich and C-rich AGB outflow as the cause.
However, since this correlation is still based on a small number of
objects, the results for a larger sample of nebulae should help clarify
whether either hypothesis fits better.

We thank the referee, Dr. R. Waters, for his useful comments. MC thanks
NASA for support under grant NAS5-4884 to UC Berkeley.

{}

\newpage

\section{Figure captions}
\noindent
\vspace*{5mm}

\begin{figcaption}
{The full SWS+LWS spectrum of CPD--56$^{\rm o}$8032 (heavy solid
line). Also shown are: its IRAS PSC fluxes (open squares, plotted at
their isophotal wavelengths); the 2-20-$\mu$m photometry of Cohen \&
Barlow (1980, bars without points; the continuum we subtracted 
(short-dashed line) and its two separate blackbody components (dotted).}
\end{figcaption}

\begin{figcaption}
{The emission from CPD--56$^{\rm o}$8032 in excess of the
two-blackbody continuum, from (a) 16--40-$\mu$m (upper); (b) 38--90-$\mu$m
(lower).  In all plots,
the heavy solid line represents the observed excess emission; the
short-dashed line, forsterite; long-dashed line, clinopyroxene;
dotted line, orthopyroxene; long-dash dotted line,
crystalline ice; light solid line, the sum of all these modeled
components; dotted line $<4~\mu$m in 1a), the stellar radiation.
The units of excess flux are W\,cm$^{-2}$\,$\mu$m$^{-1}$.}
\end{figcaption}

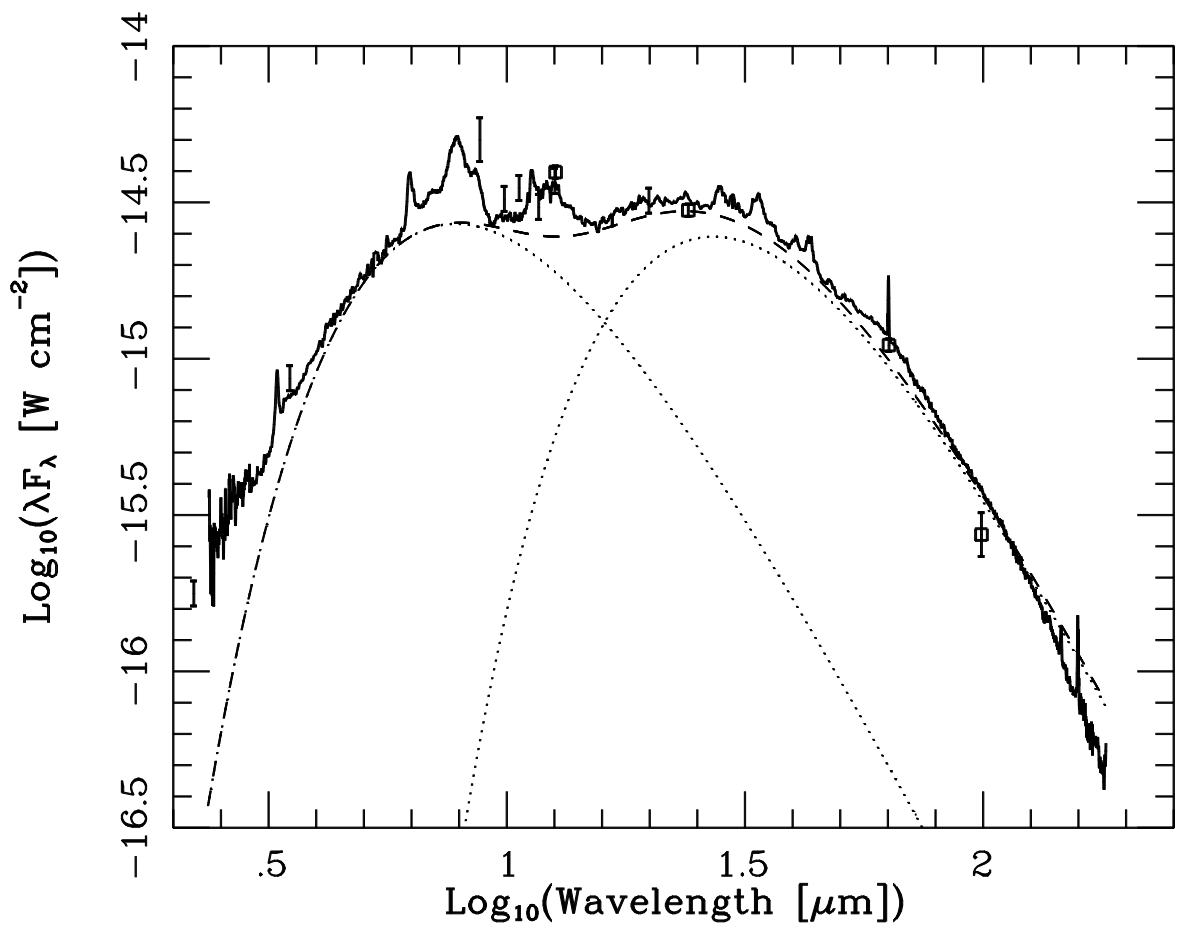
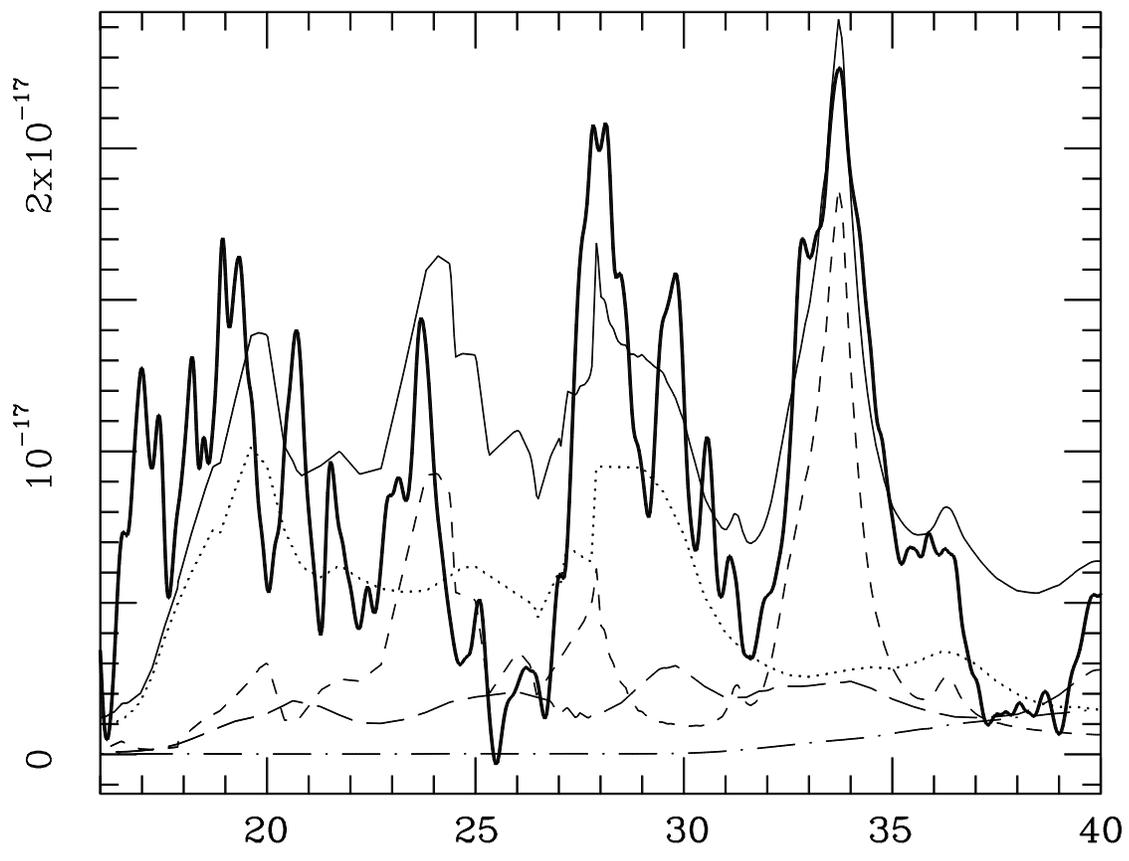
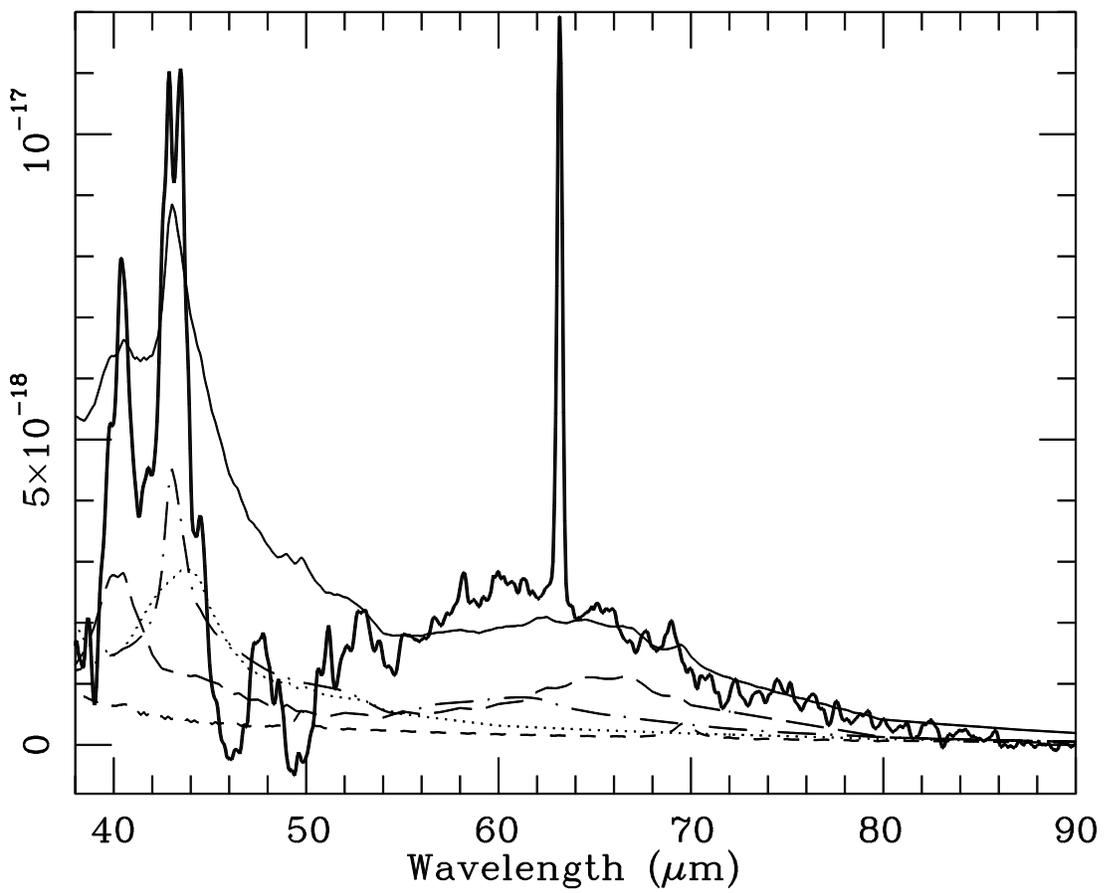
\end{document}